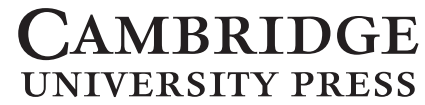

**Research Paper**

# Second-generation planet formation after tidal disruption from common envelope evolution

Luke Chamandy[1,2], Jason Nordhaus[3,4], Eric G. Blackman[2] and Emily Wilson[5]

[1]National Institute of Science Education and Research, An OCC of Homi Bhabha National Institute, Bhubaneswar 752050, Odisha, India
[2]Department of Physics and Astronomy, University of Rochester, Rochester NY 14627, USA
[3]Center for Computational Relativity and Gravitation, Rochester Institute of Technology, Rochester, NY 14623, USA
[4]National Technical Institute for the Deaf, Rochester Institute of Technology, Rochester, NY 14623, USA
[5]Department of Astronomy and Physics, Lycoming College, Williamsport, PA 17701, USA

## Abstract

We propose that certain white dwarf (WD) planets, such as WD 1856+534 b, may form out of material from a stellar companion that tidally disrupts from common envelope evolution with the WD progenitor star. The disrupted companion shreds into an accretion disc, out of which a gas giant protoplanet forms due to gravitational instability. To explore this scenario, we make use of detailed stellar evolution models consistent with WD 1856+534. The minimum mass companion that produces a gravitationally-unstable disc after tidal disruption is $\sim$0.15 $M_\odot$. In this scenario, WD 1856+534 b might have formed at or close to its present separation, in contrast to other proposed scenarios where it would have migrated in from a much larger separation. Planet formation from tidal disruption is a new channel for producing second-generation planets around WDs.

## 1. Introduction

Planets in short-period orbits around white dwarfs (WD) are predicted to be rare, as many are expected to be destroyed during post-main-sequence evolution or migrate to longer-period orbits (Nordhaus et al. 2010; Nordhaus and Spiegel 2013). However, the presence and characteristics of a planet in a short-period orbit around a WD can provide constraints on formation scenarios. First generation planets may migrate from large-period orbits via Kozai-Lidov cycles in hierarchical triple systems or near-coplanar scattering with other planets of similar mass (followed by tidal circularisation), or perhaps if they can survive a common envelope (CE) event (Lagos et al. 2021). Second generation planets have been suggested to form in the ejecta of the CE (Perets 2011; Kashi and Soker 2011; Völschow et al. 2014; Schleicher and Dreizler 2014; Bear and Soker 2014; Ledda et al. 2023). Other scenarios for second (or third) generation planet formation are discussed by Perets (2011), Perets and Kenyon (2013), and Hogg et al. (2018). Formation of second generation giant planets around isolated WD debris discs is unlikely, as the surface densities are low, but the formation of low-mass planets may be possible (Bear and Soker 2015; van Lieshout et al. 2018).

Observational searches via various methods, such as eclipses, have revealed a few interesting WDs such as WD 1856+534, estimated to have mass $M_{\rm wd} = 0.518 \pm 0.055\,{\rm M}_\odot$ (Vanderburg et al. 2020), $0.606 \pm 0.039$ (Alonso et al. 2021) or $0.576 \pm 0.040\,{\rm M}_\odot$ (Xu et al. 2021), and a detected planet, WD 1856+534 b, on a 1.41 d orbit with mass $1\,{\rm M_J} \lesssim m_2 \lesssim 12\,{\rm M_J}$ (Vanderburg et al. 2020; Xu et al. 2021). Throughout this study, we scale our results to parameter values for this system.

WD 1856+534 b is difficult to explain using a single CE scenario (Vanderburg et al. 2020; Lagos et al. 2021; Chamandy et al. 2021, hereafter C21; O'Connor et al. 2023). In C21, we proposed that WD 1856+534 b was instead dragged in from a wider orbit by a CE event involving a companion that tidally disrupts inside the envelope. In that scenario, a second CE event – the one with the observed planet – unbinds the remainder of the envelope. A somewhat different idea is that the planet was dragged in when the WD progenitor underwent a helium flash (Merlov et al. 2021). This work proposes a different scenario to explain such planets. As in C21, a CE event takes place that leads to the tidal disruption of the companion. But in our new scenario, the observed planet forms in an accretion disc that results from the tidal disruption event. Notably, a somewhat similar scenario has been proposed to explain planets orbiting millisecond pulsars. In that scenario, a WD companion is disrupted, resulting in a disc around the neutron star out of which planets form. Accretion onto the neutron star causes it to spin up to millisecond periods (e.g. Stevens et al. 1992; van den Heuvel 1992; Margalit and Metzger 2017).

Other alternatives for explaining WD 1856+534 are migration due to the Zeipel-Lidov-Kozai effect driven by the distant binary M-dwarf companion system (Muñoz and Petrovich 2020; O'Connor et al. 2021; Stephan et al. 2021) and planet-planet scattering (Maldonado et al. 2021; O'Connor et al. 2022; Maldonado et al. 2022). While the latter scenario is plausible, searches for other planets in this system have so far come up empty (Kubiak et al. 2023).

**Author for correspondence:** E.G. Blackman, Email: blackman@pas.rochester.edu

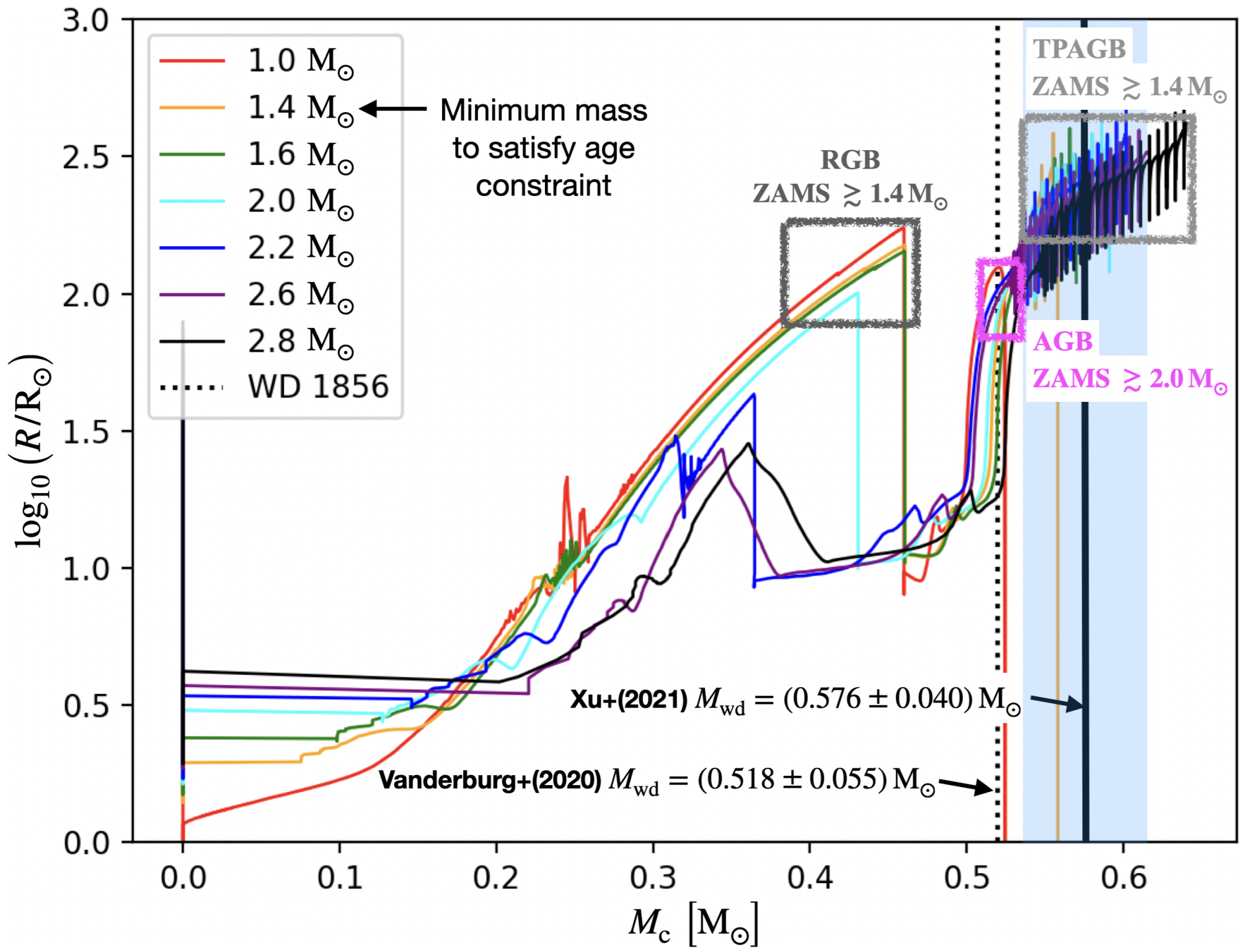

**Figure 1.** Relationship between primary radius and core mass, for different values of $M_{ZAMS}$. These results were obtained using MESA assuming solar metallicity ($Z = 0.02$) and with Reimers and Blöcker scaling coefficients set to $\eta_R = 0.7$ and $\eta_B = 0.15$, respectively, as motivated in Section 1. Constraints on $M_{ZAMS}$ for the system WD 1856+534 are highlighted in the plot. The lower limit of $\approx 1.4\,M_\odot$ comes from the age upper limit of $\sim 10\,$Gyr (Vanderburg et al. 2020). Larger states most likely to undergo CE are boxed (RGB, AGB, TPAGB). For curves enclosed by the magenta (middle) rectangle, only those with $M_{ZAMS} \gtrsim 2.0\,M_\odot$ are likely because below this the maximum radius on the RGB is greater.

Our proposed scenario begins with a CE event involving a $\sim$1.5-2.5 $M_\odot$ red giant branch (RGB) or asymptotic giant branch (AGB) star and a $\sim$0.15-1 $M_\odot$ main sequence (MS) star. In such cases, the companion tidally disrupts to form an accretion disc that orbits the core of the WD progenitor. The formation and early evolution of such discs has been studied with hydrodynamic simulations, albeit with planetary or brown dwarf rather than stellar companions (Guidarelli et al. 2019, 2022). Sufficient orbital energy can be liberated before, during, and after tidal disruption to eject the remainder of the CE, leaving a system consisting of a proto-WD core and an accretion disc of perhaps a few $\times\,0.1\,M_\odot$ in mass.

As the disc viscously spreads, it transitions from an advective, radiation pressure-dominated state with $h \sim r$ (e.g. Nordhaus et al. 2011) to a radiative cooling-dominated, gas pressure-dominated state with $h \ll r$ (e.g. Shen and Matzner 2014), where $h$ is the disc scale height at radius $r$. The disc then becomes gravitationally unstable near its outer radius and self-gravitating clumps of mass greater than the Jeans mass (of order 1 $M_J$) collapse to form puffed-up protoplanets that may be massive enough to clear a gap and avoid rapid inward (type I) migration down to their own tidal disruption separations (e.g. Boss 1997, 1998; Zhu et al. 2012; Schleicher and Dreizler 2014; Lichtenberg and Schleicher 2015).

**Table 1.** List of stellar models with parameters, obtained by running single-star simulations with the 1D stellar evolution code MESA. C18 refers to Chamandy et al. (2018). Quantities are the stellar mass $M$, mass of its zero-age MS progenitor $M_{ZAMS}$, mass of its core $M_c$, radius $R$, the value of the dimensionless parameter $\lambda$ appearing in equation (13) for the envelope binding energy $E_{bind}$, which is listed in the final column, chosen so that the binding energy includes the gravitational potential energy and thermal energy. Models were evolved using MESA release 10108 with solar metallicity ($Z = 0.02$) and with mass-loss parameters on the RGB and AGB of $\eta_R = 0.7$ and $\eta_B = 0.15$, respectively, so as to match the initial-final mass relation of Cummings et al. (2018), with the exception of the RGB(C18) model, which used MESA release 8845 with $\eta_R = 1$.

| Model | Description | $M$ | $M_{ZAMS}$ | $M_c$ | $R$ | $\lambda$ | $E_{bind}$ |
|---|---|---|---|---|---|---|---|
| | | $[M_\odot]$ | $[M_\odot]$ | $[M_\odot]$ | $[R_\odot]$ | | $[10^{47}$ erg$]$ |
| A | RGB(C18) | 2.0 | 2.0 | 0.37 | 48 | 1.3 | 1.9 |
| B | RGB | 1.5 | 1.6 | 0.40 | 85 | 0.7 | 1.1 |
| C | RGB($\sim$tip) | 1.4 | 1.6 | 0.46 | 140 | 0.6 | 0.6 |
| D | AGB | 2.2 | 2.2 | 0.52 | 100 | 0.7 | 2.0 |
| E | TPAGB(lowM) | 1.4 | 1.6 | 0.55 | 250 | 0.4 | 0.5 |
| F | TPAGB(highM) | 2.0 | 2.2 | 0.57 | 250 | 0.7 | 0.6 |

## 2. Constraining the proposed scenario

### 2.1. Progenitor

To obtain realistic examples of RGB and AGB progenitors, we employ Modules for Experiments in Stellar Astrophysics (MESA) (Paxton et al. 2011, 2013, 2015, 2019). Table 1 summarizes the progenitor models we consider. All models assume solar metallicity ($Z = 0.02$). Models B-F are consistent with the observed initial-final mass relationships derived from cluster observations (Cummings et al. 2018; Hollands et al. 2023) and have been extensively used for CE studies (Wilson and Nordhaus 2019, 2020, 2022; Kastner and Wilson 2021). Model A has been used to model the initial stellar profile in a series of papers involving 3D simulations, starting with Chamandy et al. (2018), and uses the same parameter values as the other models except for a slightly different value of the Reimers mass loss parameter ($\eta_{\rm R} = 1$ instead of 0.7).

Figure 1 shows the stellar radius as a function of core mass for stellar models of various zero-age main sequence (ZAMS) masses and parameter values as in Models B-F. The core mass increases with time, so stars evolve from left to right. Observational mass estimates for WD 1856+534 are plotted as vertical lines for reference. That the observed WD mass coincides with the AGB phase suggests that the progenitor was on the AGB at the time of the CE event. However, as discussed below, the proto-WD core mass can increase by accreting the disrupted companion, so an RGB progenitor cannot immediately be ruled out.

Given that the age of the WD 1856+534 system is $\lesssim 10$ Gyr and the cooling age of the WD is $5.85 \pm 0.5$ Gyr (Vanderburg et al. 2020), the WD progenitor spent $\lesssim 4.6$ Gyr on the MS, which implies a mass of $M_{\rm ZAMS} \gtrsim 1.4\,{\rm M_\odot}$.[a] Furthermore, an AGB progenitor would likely have had $M_{\rm ZAMS} \gtrsim 2.0\,{\rm M_\odot}$; otherwise its RGB radius would have been larger and it would have entered CE then. Given the steepness of the initial mass function above $1\,{\rm M_\odot}$ (Salpeter 1955; Hennebelle and Grudić 2024), we choose examples with $M_{\rm ZAMS} \leq 2.2\,{\rm M_\odot}$ (Table 1), but more massive progenitors are possible.

### 2.2. Disc formation and survival

A fraction of the disrupted companion forms an accretion disc around the AGB or RGB core, inside the remaining common envelope (Reyes-Ruiz and López 1999; Blackman et al. 2001; Nordhaus and Blackman 2006; Nordhaus et al. 2011; Nordhaus and Spiegel 2013; Guidarelli et al. 2019, 2022). The hydrodynamic simulations performed by Guidarelli et al. (2022) involved 10-30 $\rm M_J$ companions undergoing tidal disruption inside the envelope of an AGB star. About 60% of the mass of the disrupted companion formed a disc and the other 40% constituted an outwardly moving but gravitationally bound tidal tail that would eventually fall back. This is reminiscent of work by Shen and Matzner (2014) on tidal disruption event discs around supermassive black holes. These authors find that a large fraction of the material from the disruption falls back and collides with itself, settling at a radius somewhat larger than the disruption radius. Guidarelli et al. (2022) find that the disc becomes quasi-Keplerian within a few dynamical times, with an aspect ratio $0.05 \lesssim h/r \lesssim 0.2$. Metzger et al. (2021) models the tidal disruption of a star in a cataclysmic variable system and the disc that subsequently forms around the WD; the model explored in this work is in some ways similar to their model.

Guidarelli et al. (2019) simulated a disc surrounded by a hot AGB envelope, and estimated that the disc is likely to be stable for at least 10 times the duration of the simulation, or 100 orbits at the outer radius of the disc. For a disc of outer radius equal to the minimum disruption separation $\sim 0.3\,{\rm R_\odot}$ (Figure 2a and Section 2.5), this corresponds to $\sim 3$ d, whereas taking the disc outer radius to be the present orbital separation of WD 1856+534b of $4.4\,{\rm R_\odot}$ gives $\sim 0.4$ yr. The actual disc survival time might be much longer, especially considering that the envelope would have expanded and spun up due to orbital energy and angular momentum transfer from the companion, reducing the destabilizing effect of shear on the disc.

If the envelope were to truncate the evolution of the disc and outlive it, this could provide challenges for the formation, survival, and emergence of the planet in its current orbit. However, we find that the envelope is likely removed before, during, or shortly after disc formation. In Appendix E, we discuss what processes may power envelope ejection, and estimate how long it takes. One possibility is that tidal disruption of the companion is preceded by a Roche lobe overflow (RLOF) phase that leads to unstable mass transfer and orbital tightening, and the transfer of orbital energy ejects the envelope prior to disruption. In this case, the disc of disrupted companion material would not be surrounded by a hot envelope, so would easily survive. Another possibility is that the envelope is removed by the release of energy during an early, possibly brief phase of advection-dominated accretion. We speculate that envelope unbinding might also be assisted by shocking during disc formation and nuclear reactions due to accretion onto the core.

If the envelope is not rapidly unbound by these processes, then it could be unbound by subsequent accretion at close to the Eddington rate, but this takes $\sim 10^2$ years. Whether the disc can survive for this long is not clear. In cases where early rapid removal of the envelope does not occur, survivability of the disc may be a bottleneck in our planet formation scenario that may help to explain why such planets are rare. In any case, after the envelope is ejected, ionizing radiation from the hot central core photo-evaporates the disc in $\sim 10$ Myr (Appendix A).

### 2.3. Minimum required companion and protoplanet masses

From the Toomre criterion (Toomre 1964), the disc is gravitationally unstable if

$$Q = \frac{h\Omega^2}{\pi G \Sigma} \lesssim 1, \tag{1}$$

where

$$\Omega = \left(\frac{GM_{\rm c}}{r^3}\right)^{1/2} \tag{2}$$

is the angular rotation speed with $M_{\rm c}$ the mass of the core of the giant star, and $\Sigma$ is the gas surface density. We approximate the surface density as (e.g. Armitage and Rice 2005; Raymond and Morbidelli 2022)

$$\Sigma = \Sigma_{\rm out}\left(\frac{r}{r_{\rm out}}\right)^{-\beta}, \tag{3}$$

with $1 \lesssim \beta \lesssim 3/2$. A fraction $f$ of the disrupted companion forms the disc, while the rest either accretes prior to disc formation or mixes with envelope material.

In Appendix B, we show that condition (1) leads to the following constraint on the disc mass:

$$f m_1 \gtrsim 0.12\,{\rm M_\odot}\frac{1}{2-\beta}\left(\frac{\theta}{0.1}\right)\left(\frac{M_{\rm c}}{0.576\,{\rm M_\odot}}\right)\left(\frac{r}{r_{\rm out}}\right)^{-(2-\beta)}. \tag{4}$$

[a]Using the estimate $t_{\rm cool} = 6.60 \pm 0.48$ Gyr of Xu et al. 2021 would slightly increase this lower limit

Since $f \leq 1$, gravitational instability requires $m_1 \gtrsim 0.12\,\mathrm{M_\odot}$, with $m_1$ the mass of the disrupted companion. (Throughout, we use subscript '1' to refer to the disrupted companion and subscript '2' to refer to the extant companion, i.e. the planet.) For equation (2), we neglected the disc self-gravity but it can be important if $fm_1$ is comparable to $M_\mathrm{c}$.

The mass of the collapsing object must also exceed the Jeans mass (e.g. Binney and Tremaine 2008),

$$M_\mathrm{Jeans} = \frac{\pi^{5/2} c_\mathrm{s}^3}{6G^{3/2}\rho^{1/2}}, \tag{5}$$

where $c_\mathrm{s}$ is the sound speed. Substituting $c_\mathrm{s} = h\Omega$, $\rho = \Sigma/2h$, $h = \theta r$, $\Omega$ from equation (2), and $\Sigma$ from equations (3) and (22), we obtain

$$M_\mathrm{Jeans} \approx 2.7\,\mathrm{M_J} \frac{1}{(2-\beta)^{1/2}} \left(\frac{M_\mathrm{c}}{0.576\,\mathrm{M_\odot}}\right)^{3/2} \left(\frac{fm_1}{0.3\,\mathrm{M_\odot}}\right)^{-1/2} \times \left(\frac{\theta}{0.1}\right)^{7/2} \left(\frac{r}{r_\mathrm{out}}\right)^{-1+\beta/2}. \tag{6}$$

If WD 1856+534 b formed this way, all of the material in the collapsing protoplanet was incorporated into the planet, and the planet retained the same mass up to the present time, then this would imply that $m_2 \gtrsim M_\mathrm{Jeans}$, but some protoplanetary mass could be lost due to inefficiencies or ablation in the formation process. Regardless, this result is consistent with the observational estimate, $0.84 \lesssim m_2/\,\mathrm{M_J} \lesssim 11.7$ (Vanderburg et al. 2020; Xu et al. 2021), as well as with the lower limit of $2.4\,\mathrm{M_J}$ found by Alonso et al. (2021).

## 2.4. Timescale for planet formation

The protoplanet contracts on the free-fall timescale, $t_\mathrm{ff} \lesssim 1$ d (see Appendix C). This can be compared with the viscous dissipation timescale

$$t_\mathrm{visc} = \frac{r^2}{\nu} \approx \frac{r^2}{\alpha_\mathrm{SS} c_\mathrm{s} h} \approx \frac{r^2}{\alpha_\mathrm{SS} h^2 \Omega}, \tag{7}$$

where $\alpha_\mathrm{SS}$ is the viscosity parameter (Shakura and Sunyaev 1973). Making use of equation (2) we obtain

$$t_\mathrm{visc} \approx 6\,\mathrm{yr} \left(\frac{r}{4.4\,\mathrm{R_\odot}}\right)^{3/2} \left(\frac{\theta}{0.1}\right)^{-2} \left(\frac{M_\mathrm{wd}}{0.576\,\mathrm{M_\odot}}\right)^{-1/2} \left(\frac{\alpha_\mathrm{SS}}{0.01}\right)^{-1}, \tag{8}$$

which easily exceeds $t_\mathrm{ff}$, facilitating giant gaseous protoplanet (GGPP) by direct collapse (Boss 1998). Transformation of the GGPP into a full-fledged planet takes much longer, and is mostly independent of the disc evolution other than possible migration.

## 2.5. Companion mass upper limit and the WD progenitor

The disrupted companion must have been massive enough for the disc to satisfy condition (4); but not so massive as to have avoided disruption by unbinding the envelope. The orbital separation at which tidal disruption occurs is given by (e.g. Nordhaus and Blackman 2006)

$$a_\mathrm{d1} \approx \left(\frac{2M_\mathrm{c}}{m_1}\right)^{1/3} r_1, \tag{9}$$

where $M_\mathrm{c}$ is the mass of the core of the giant and $r_1$ is the radius of the companion. Figure 2a shows $a_\mathrm{d1}$ as a function of $m_1$ (solid lines). The colours represent stellar models listed in Table 1. Dotted lines show the orbital separation at which RLOF occurs (Eggleton 1983),

$$a_{\rm RLOF} = \frac{0.6q^{2/3} + \ln(1+q^{1/3})}{0.49q^{2/3}} r_1, \quad (10)$$

where $q = M_c/m_1$. If $a_{\rm RLOF} > a_{\rm d1}$, then the companion begins losing mass at around $a_{\rm RLOF}$. Assuming this mass transfer phase is unstable, the separation continues to decrease and the companion tidally disrupts at around $a_{\rm d1}$. If $a_{\rm RLOF} < a_{\rm d1}$, then tidal disruption happens directly, without any RLOF phase. To estimate the radius of the disrupted companion and that of the planet, we use the following approximation to the results of Chabrier et al. (2009):

$$\frac{r}{\rm R_\odot} = \begin{cases} 0.1 & \text{if } m/{\rm M_\odot} \le 0.077; \\ 0.1\left(\dfrac{m/{\rm M_\odot}}{0.077}\right)^{0.8} & \text{if } 0.077 < m/{\rm M_\odot} \le 0.4; \\ 0.1\left(\dfrac{0.4}{0.077}\right)^{0.8}\left(\dfrac{m/{\rm M_\odot}}{0.4}\right)^{1.075} & \text{if } m/{\rm M_\odot} > 0.4. \end{cases} \quad (11)$$

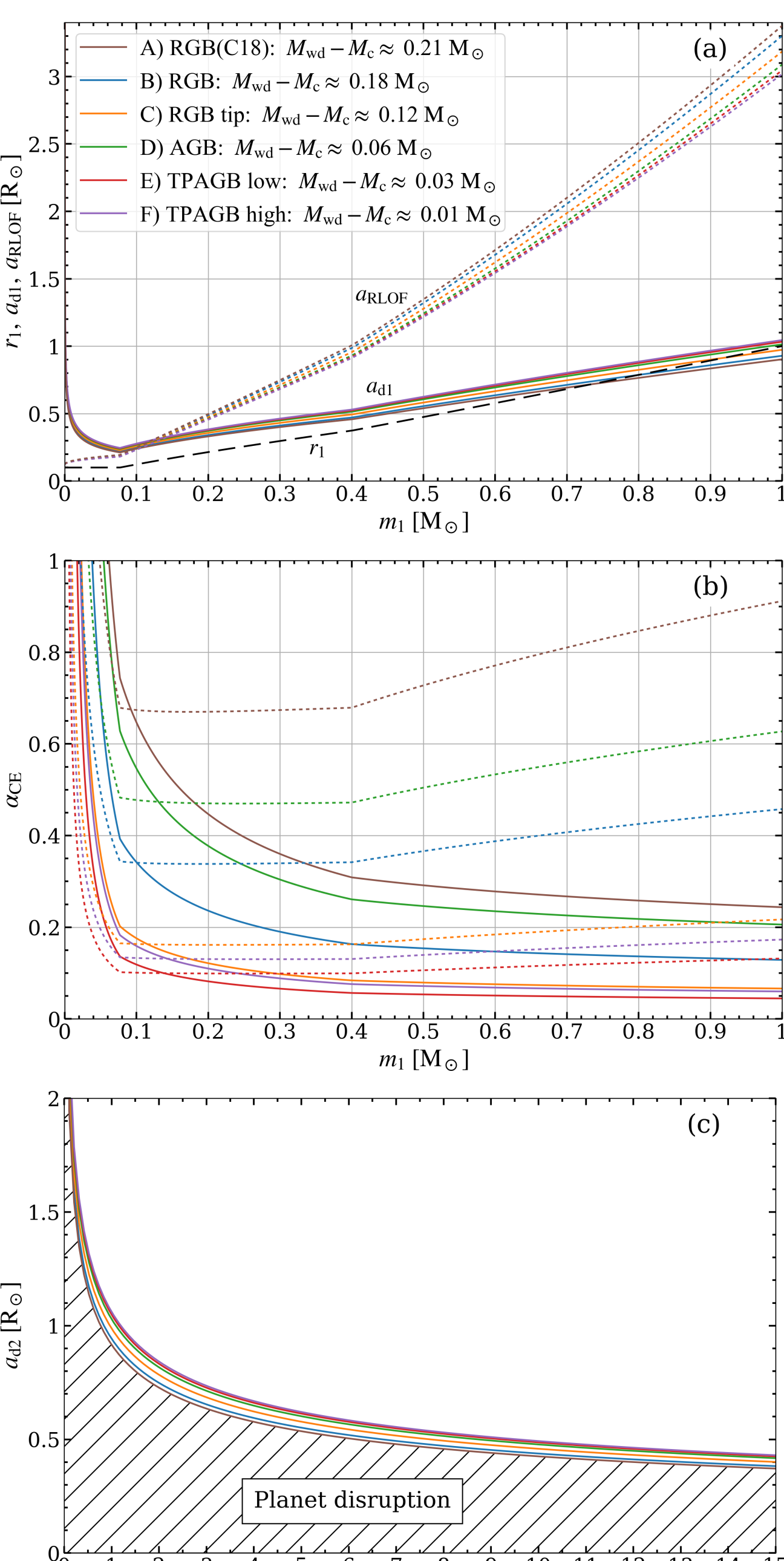

**Figure 2. Panel (a):** Orbital separation where stellar tidal disruption occurs $a_\mathrm{d1}$ (solid lines), and where RLOF is initiated if the stellar companion has not already disrupted (dotted lines) for the giant star models of Table 1, as a function of the companion mass. The companion radius is shown as a dashed line (equation 11) and the difference between the observed WD mass according to Xu et al. (2021) and the core mass in the stellar model is shown in the legend. **Panel (b):** Maximum allowed value of the common envelope efficiency parameter $\alpha_\mathrm{CE}$ as a function of the companion mass; above this value the companion unbinds the envelope before it can be disrupted (solid lines) or before RLOF is initiated (dotted lines). It is possible that the envelope could be ejected *before* disruption if RLOF leads to unstable mass transfer and inspiral down to $a_\mathrm{d1}$. Thus, for a given progenitor, the parameter space above the dotted line is excluded, and that below the dotted line but above the solid line is viable only if RLOF is initiated and ultimately leads to disruption. **Panel (c):** As the top panel, but now zoomed in to show the relevant parameter space for the *planet*, and with mass shown in units of $\mathrm{M_J}$ ($1\,\mathrm{M_J} \approx 10^{-3}\,\mathrm{M_\odot}$). The planet cannot be formed in the hatched region, which corresponds to separations less than its tidal disruption separation $a_\mathrm{d2}$.

The value of $r_1$ is shown as a black dashed line in Figure 2a. It can be seen that $r_1 > a_{\rm d1}$ in some cases for large values of $m_1$, which means the model is not fully self-consistent for that small region of parameter space. This limitation of the model stems from its simplicity.[b] However, as such lack of precision in the modelling does not affect our final conclusions, we choose to keep the model as simple as possible.

The envelope remains bound if the liberated orbital energy $\Delta E_{\rm orb}$ multiplied by the efficiency parameter $\alpha_{\rm CE}$ is smaller than the envelope binding energy $E_{\rm B}$, i.e.,

$$\alpha_{\rm CE}\Delta E_{\rm orb} < E_{\rm B}, \quad (12)$$

where

$$E_{\rm B} = \frac{GM(M-M_c)}{\lambda R}, \quad (13)$$

$M$ and $R$ are the mass and radius of the WD progenitor, and $\lambda$ is a parameter of order unity that depends on the stellar density and temperature profiles. To satisfy constraint (12) at the tidal disruption separation, we substitute $\Delta E_{\rm orb} \approx GM_c m_1/2a_{\rm d1}$, where we have neglected the initial orbital energy. To satisfy the same constraint at the separation at which RLOF begins, we instead substitute $\Delta E_{\rm orb} \approx GM_c m_1/2a_{\rm RLOF}$.

Figure 2b shows the upper limit to $\alpha_{\rm CE}$ below which the envelope is not completely unbound before the tidal disruption separation is reached (solid lines). The value of $\alpha_{\rm CE}$ is not well constrained and likely varies from one system to another, but estimates for low-mass systems are typically between 0.05 and 0.5 (e.g. Iaconi and De Marco 2019; Wilson and Nordhaus 2019; Scherbak and Fuller 2023). The same lines can be used to find the maximum value of $m_1$ below which the envelope remains bound, for a given value of $\alpha_{\rm CE}$. However, RLOF inside the envelope may lead to runaway mass loss, orbital tightening, and tidal disruption, even if the envelope were ejected shortly after the onset of RLOF. The dotted lines again show upper limits on $\alpha_{\rm CE}$, but now requiring that the envelope remains bound until the onset of RLOF, which occurs at the separation $a_{\rm RLOF}$. For $m_1 \gtrsim 0.1\,{\rm M_\odot}$, we see from Figure 2a that $a_{\rm RLOF} > a_{\rm d1}$, so RLOF could occur for $m_1 \gtrsim 0.1\,{\rm M_\odot}$.

Figure 2b shows that the more distended models (RGB tip, shown in orange, and high/low mass thermally pulsing AGB (TPAGB), shown in violet/red) require $m_1 \lesssim 0.08\,{\rm M_\odot}$ if $\alpha_{\rm CE} = 0.2$, using the more conservative limit (solid lines). If $\alpha_{\rm CE} = 0.1$, then the maximum value of $m_1$ lies within the range $(0.13, 0.28)$ for these three models. Models with $\alpha_{\rm CE} \ge 0.2$ are ruled out because condition (4) is not satisfied: the orange, violet and red lines all intersect $\alpha_{\rm CE}$ at $m_1 < 0.1\,{\rm M_\odot}$. If $0.1 < \alpha_{\rm CE} < 0.2$, the allowed range of $m_1$ is quite narrow. By constrast, the less distended RGB and AGB models accommodate a larger range of companion mass because the magnitude of the envelope binding energy is larger. Focussing, conservatively, on the solid rather than the dotted lines, the AGB model (Model D, green) admits companion masses up to $\approx 1\,{\rm M_\odot}$ if $\alpha_{\rm CE} = 0.2$, up to $\approx 0.3\,{\rm M_\odot}$ if $\alpha_{\rm CE} = 0.3$, and up to $\approx 0.18\,{\rm M_\odot}$ if $\alpha_{\rm CE} = 0.4$. The RGB(C18) model (Model A, brown) allows for slightly larger upper limits to the companion mass.

CE evolution culminating in tidal disruption of a companion massive enough to form a planet is thus more likely for *less distended* primary stars. Using the less conservative upper limit on $\alpha_{\rm CE}$ corresponding to RLOF (dotted lines), the allowed ranges of $\alpha_{\rm CE}$ and $m_1$ are larger, but less distended progenitors are still favoured.

### 2.6. Implications of accretion for the WD mass

In the legend of Figure 2, we show the approximate difference between $M_{\rm wd} \approx 0.576\,{\rm M_\odot}$ (Xu et al. 2021) and the core mass of the primary (Table 1). This difference could be explained by accretion of material from the disrupted companion. For the RGB models, $\sim$0.12-0.21 ${\rm M_\odot}$ must be accreted, whereas for the AGB/TPAGB models only $\sim$0.01-0.06 ${\rm M_\odot}$ must be accreted. The AGB model (Model D, green) and RGB(C18) model (Model A, brown) can accommodate a larger range of disrupted companion masses than the other models (Section 2.5), but the AGB model requires less accretion of disrupted companion material onto the proto-WD core and may thus be more likely. In Appendix D we estimate that $\sim 10^{-3}$-$10^{-2}\,{\rm M_\odot}$ is accreted during a phase of advection-dominated accretion, but that another $\sim 0.1\,{\rm M_\odot}$ may accrete in the next $10^3$-$10^4$ yr during an Eddington-limited phase.

### 2.7. Location of planet formation

The planet must form outside of its own tidal disruption separation $a_{\rm d2}$, plotted in Figure 2c as a function of the planet mass $m_2$. For $1 < m_2/{\rm M_J} < 12$, the range is $0.5 \lesssim a_{\rm d2} \lesssim 1.0\,{\rm R_\odot}$, but $a_{\rm d1} < 0.5\,{\rm R_\odot}$ for $m_1 \lesssim 0.4\,{\rm M_\odot}$, and $a_{\rm d1} < 1\,{\rm R_\odot}$ for $m_1 \lesssim 1.0\,{\rm M_\odot}$, as shown in Figure 2a, ignoring $m_1 < 0.01\,{\rm M_\odot}$, which is excluded by the lower limit (4). Thus, if the planet formed at $a_{\rm d1}$, this would imply a separate lower limit for $m_1$ in the range 0.5-1.0 ${\rm M_\odot}$ due to the minimum in the $a_{\rm d1}$ vs. mass relation plotted in Figure 2a. But could the planet instead form at a larger separation, or, for WD 1856+534 b, at its present location $a_2 \approx 4.4\,{\rm R_\odot}$?

For $\theta = 0.1$, the disc can viscously spread out to to $r_{\rm out} \sim a_2$ on the viscous timescale, $\sim$6 yr (equation 8). But if the flow is advection-dominated then $\theta \sim 1$, lowering the diffusion time by a factor of $\sim$100. At some point the disc transitions from the advection-dominated geometrically thick regime to the gas-pressure-dominated geometrically thin regime (e.g. Shen and Matzner 2014). Protoplanet formation can occur once $Q$ drops below unity and $Q$ is proportional to $\theta$ and to a negative power of $r/r_{\rm out}$ (equation 25). Thus, when the disc transitions, $\theta$ and $Q$ drop by an order of magnitude so this transition might trigger gravitational instability in the outer disc, leading to protoplanet formation at separations $> a_{\rm d1}$. Thus, WD 1856+534 b might have formed at or close to its present separation.

[b] For example, assuming synchronous rotation of the companion and taking into account the centrifugal force (but still ignoring tidal deformation) increases $a_{\rm d1}$ by the factor $(3/2)^{1/3} \approx 1.145$, which is sufficient to rectify this inconsistency for the values of $m_1$ plotted.

By the end of the simulations of Guidarelli et al. (2022), the expanding tidal tail formed during the disruption extends an order of magnitude larger than the tidal disruption separation $a_{\rm d1}$. This material may fall back and extend the disc (e.g. Shen and Matzner 2014). Mass transfer and disc formation may also begin in the RLOF phase, which begins at separations $> a_{\rm d1}$ (Figure 2a). Moreover, the companion likely inflates by accreting a quasi-hydrostatic atmosphere (Chamandy et al. 2018), which would, in principle, cause it to overflow its Roche lobe at still larger separation. RLOF might also lead to mass transfer through the L2 point and circumbinary disc formation (e.g. MacLeod et al. 2018). These processes could increase the outer radius of the disc, facilitating planet formation at separations $> a_{\rm d1}$.

### 2.8. Minimum mass to prevent type I migration

Once formed, the protoplanet can avoid destruction due to type I inward migration by opening a gap if (e.g. Papaloizou 2021)

$$m_2 \gtrsim 2\,{\rm M_J} \left(\frac{M_{\rm wd}}{0.576\,{\rm M_\odot}}\right) \left(\frac{\theta}{0.1}\right)^3 \tag{14}$$

and

$$m_2 \gtrsim 2\,{\rm M_J} \left(\frac{M_{\rm wd}}{0.576\,{\rm M_\odot}}\right) \left(\frac{\theta}{0.1}\right)^2 \left(\frac{\alpha_{\rm SS}}{0.01}\right), \tag{15}$$

which happen to be lower limits of similar magnitude to each other and to $M_{\rm Jeans}$ (equation 6). Therefore, two separate lines of reasoning favour a planet mass $\gtrsim 2\,{\rm M_J}$; this lower limit is thus a fairly robust prediction of our model.

## 3. Summary and Conclusions

We have presented a second generation planet formation scenario that may explain the occurrence of giant planets in low-period orbits around WDs. In this scenario, planets form from gravitational instability of an accretion disc formed out of material from a low-mass star that was tidally disrupted due to CE interaction with the WD progenitor core. We showed that WD 1856+534 b may be explained by this scenario if: (i) the WD progenitor was a relatively compact AGB star with $M \gtrsim 2\,{\rm M_\odot}$, (ii) the mass of the disrupted companion was in the range $0.15 \lesssim m_1/{\rm M_\odot} \lesssim 1\,{\rm M_\odot}$, and (iii) the mass of the observed planet $m_2 \gtrsim 2\,{\rm M_J}$. The latter can be considered as a prediction for WD 1856+534 b, and is similar to the lower limits obtained using transmission spectroscopy of $2.4\,{\rm M_J}$ (Alonso et al. 2021) and $0.84\,{\rm M_J}$ (Xu et al. 2021), while remaining comfortably below the upper limit of $\sim 12\,{\rm M_J}$, which is based on the non-detection of thermal emission (Vanderburg et al. 2020).

WD 1856+534 b might have formed at the orbital separation inferred from observations or might have migrated to it. Detailed models (e.g. Masset and Papaloizou 2003) show that the sense and timescale of type II migration can vary in time and can be sensitive to various parameters. The interplay between migration and evaporation by the hot white dwarf core of the primary may account for the rarity of systems like WD1856+534 b, depending on where the planet forms before migration and how long it spends close enough to the WD core to be evaporated (Gänsicke et al. 2019; Schreiber et al. 2019; Lagos et al. 2021; Gallo et al. 2024). Future work is needed to address these issues.

*Acknowledgements.* We are grateful to the first referee for comments that helped to improve the presentation of this work, and to the second referee for various helpful comments. L.C. thanks R.I.T. for supporting his visit from 4-10 June, 2023.

*Funding Statement.* Financial support for this project was provided by the U. S. Department of Energy grants DE-SC0001063, DE-SC0020432 and DE-SC0020103, the U. S. National Science Foundation grants AST-1515648, AST-1813298, PHY-2020249, AST-2009713, and AST-2319326, and the Space Telescope Science Institute grant HST-AR-12832.01-A.

*Competing Interests.* None

*Data Availability Statement.* The codes used to produce the various figures are available upon request.

## Appendix A. Photo-evaporation timescale

The ionizing luminosity is sensitive to the effective temperature. The hottest white dwarfs are observed to have $T_{\rm eff} \sim 10^5$ K (Bédard et al. 2017). Theoretical models which track WD properties as a function of time predict that they are born with $T_{\rm eff} \approx 0.9$-$1.1 \times 10^5$ K and radius $R_{\rm wd} \approx 0.022$-$0.027\,{\rm R_\odot}$ for $M_{\rm wd} \sim 0.55$-$0.60\,{\rm M_\odot}$ (Bédard et al. 2020).[c] Or perhaps the exposed core would resemble a hot subdwarf B (sdB) star. Such stars are remnants of evolved stars often found in post-CE binary systems, and have $T_{\rm eff} \approx 2$-$7 \times 10^4$ K (Heber 2016; Ge et al. 2024).

For $T_{\rm eff} = 10^5$ K, the Planck spectrum peaks at 29 nm, in the extreme ultraviolet. Let us assume, conservatively, that all photons are ionizing. The radiative flux impinging on the disc can then be equated with the wind power, with an efficiency factor $\varepsilon$,

$$4\pi R_{\rm c}^2 \sigma T_{\rm eff}^4 \frac{\theta}{\pi/2} \varepsilon \approx \frac{1}{2} \dot{M}_{\rm w} \frac{2GM_{\rm c}}{r}, \tag{16}$$

where $\theta = h/r$ is the disc aspect ratio (assumed to be small), and we have used the escape speed from the central core for the wind speed. This gives a wind mass-loss rate of

$$\begin{aligned}\dot{M}_{\rm w} \approx 4 \times 10^{-8}\,{\rm M_\odot\,yr^{-1}} \left(\frac{\varepsilon}{0.1}\right)\left(\frac{\theta}{0.1}\right)\left(\frac{R_{\rm c}}{0.025\,{\rm R_\odot}}\right)^2 \\ \times \left(\frac{T_{\rm eff}}{10^5\,{\rm K}}\right)^4 \left(\frac{r}{4.4\,{\rm R_\odot}}\right)\left(\frac{M_{\rm c}}{0.576\,{\rm M_\odot}}\right)^{-1},\end{aligned} \tag{17}$$

where we have scaled $r$ to the present orbital separation of WD 1856+534 b, assuming a circular orbit. If we instead adopt typical values for sdB stars, with $R_{\rm c}$ about 5 times higher and $T_{\rm eff}$ about 2 times lower than the above values, we obtain approximately the same numerical estimate for $\dot{M}_{\rm w}$.

Alternatively, we can try to apply detailed models from the literature which were designed for classical protoplanetary discs. We try the model of Alexander et al. (2006) (see also Alexander et al. 2014 and Kunitomo et al. 2020), which takes as input the number of ionizing photons emanating from the star per unit time $\Phi$. We estimate

$$\Phi \approx \frac{4\pi R_{\rm c}^2 \sigma T_{\rm eff}^4}{h\nu_{\rm max}}, \tag{18}$$

with $\nu_{\rm max}$ given by Wien's displacement law. Thus, we obtain

$$\Phi \approx 6 \times 10^{45}\,{\rm s^{-1}} \left(\frac{R_{\rm c}}{0.025\,{\rm R_\odot}}\right)^2 \left(\frac{T_{\rm eff}}{10^5\,{\rm K}}\right)^3. \tag{19}$$

Then, using the Alexander et al. (2006) model with $CD = 1$, $a = 6$ and $\mu = 1$, and taking the outer disc radius to be much larger than the inner disc radius,

[c]See https://www.astro.umontreal.ca/~bergeron/CoolingModels/.

we find

$$\dot{M}_{\rm w} \approx 6 \times 10^{-9}\,{\rm M_\odot\,yr^{-1}} \left(\frac{\theta}{0.1}\right)^{-1/2} \left(\frac{\Phi}{6\times 10^{45}\,{\rm s^{-1}}}\right)^{1/2} \times \left(\frac{r_{\rm in}}{4.4\,{\rm R_\odot}}\right)^{1/2}, \tag{20}$$

where $r_{\rm in}$ is the inner radius of the disc. Thus, this estimate gives a mass-loss rate that is of the same order of magnitude as that obtained in equation (17). A disc of mass $0.3\,{\rm M_\odot}$ would take $\sim$10 Myr to evaporate, which is long compared to the timescales of other key processes, as discussed below.

## Appendix B. Gravitational instability of the disc

The disc mass can be obtained by integrating equation (3), which gives

$$fm_1 = 2\pi\Sigma_{\rm out} r_{\rm out}^{\beta} \int_{r_{\rm in}}^{r_{\rm out}} r^{1-\beta}\,dr \approx \frac{2\pi\Sigma_{\rm out} r_{\rm out}^2}{2-\beta}, \tag{21}$$

where $f$ is the fraction of the disrupted companion mass incorporated in the disc and we have assumed $\beta < 2$ and $(r_{\rm in}/r_{\rm out})^{2-\beta} \ll 1$. Rearranging, we obtain

$$\Sigma_{\rm out} \approx \frac{(2-\beta) fm_1}{2\pi r_{\rm out}^2}. \tag{22}$$

The volume density of the disc is given by

$$\rho = \frac{\Sigma}{2h} = \rho_{\rm out}\left(\frac{r}{r_{\rm out}}\right)^{-(\beta+1)}, \tag{23}$$

where $\rho_{\rm out} = \Sigma_{\rm out}/(2\theta r_{\rm out})$ with $\theta \equiv h/r$ the disc aspect ratio. Substituting expression (22) for $\Sigma_{\rm out}$ into this expression for $\rho_{\rm out}$ and then substituting into equation (23) gives

$$\rho \approx \frac{(2-\beta) fm_1}{4\pi r_{\rm out}^3 \theta}\left(\frac{r}{r_{\rm out}}\right)^{-(\beta+1)}. \tag{24}$$

A disc with $\theta = 0.1$, $fm_1 = 0.08\,{\rm M_\odot}$ and $r_{\rm out} = a_2 \approx 4.4\,{\rm R_\odot}$ has $\rho(r_{\rm out}) \approx 4 \times 10^{-3}(2-\beta)\,{\rm g\,cm^{-3}}$. If $r_{\rm out} = 100\,{\rm R_\odot}$, then $\rho(4.4\,{\rm R_\odot}) \approx 2\times 10^{-4}\,{\rm g\,cm^{-3}}$ if $\beta = 1$ or $\approx 5\times 10^{-4}\,{\rm g\,cm^{-3}}$ if $\beta = 3/2$. On the other hand, the density of the envelope of a ZAMS $2\,{\rm M_\odot}$ AGB star is $\rho_{\rm e} \approx 10^{-4}\,{\rm g\,cm^{-3}}$ at $r = 4.4\,{\rm R_\odot}$. Thus, the disc is expected to have slightly higher density than the original envelope at the present orbital separation of WD 1856+534 b. However, at this stage, the envelope would have already experienced expansion and at least partial ejection, so $\rho_{\rm e}$ would be significantly smaller than the above estimate and thus much smaller than the disc density.[d]

Combining equations (1), (2), (3) and (22), we obtain

$$Q \approx \frac{2\theta}{2-\beta}\frac{M_{\rm c}}{fm_1}\left(\frac{r}{r_{\rm out}}\right)^{-(2-\beta)}. \tag{25}$$

Thus, $Q$ decreases with $r$, and reaches a minimum at the disc outer radius $r_{\rm out}$. Now we can impose the condition for instability $Q \lesssim 1$ (constraint 1), which leads to constraint (4) on the mass of the disrupted companion.

## Appendix C. Free-fall timescale for planet formation

The free-fall timescale is given by

$$t_{\rm ff} = \left(\frac{3\pi}{32 G\rho}\right)^{1/2}. \tag{26}$$

Making use of equation (24) we obtain

$$t_{\rm ff} \approx \frac{0.2\,{\rm d}}{(2-\beta)^{1/2}}\left(\frac{\theta}{0.1}\right)^{1/2}\left(\frac{r_{\rm out}}{4.4\,{\rm R_\odot}}\right)^{3/2}\left(\frac{fm_1}{0.3\,{\rm M_\odot}}\right)^{-1/2} \times \left(\frac{r}{r_{\rm out}}\right)^{(\beta+1)/2}, \tag{27}$$

[d]To get an idea of how fast the envelope density near the companion can decrease relative to the initial value at that radius in the envelope, see, e.g., Ricker and Taam (2012), Iaconi et al. (2017), and Chamandy et al. (2019).

which is somewhat smaller than the observed orbital period of 1.41 d for WD 1856+534 b (Vanderburg et al. 2020).

## Appendix D. Accretion onto the proto-WD

At early times, accretion of disc material onto the core of the primary may occur on the viscous timescale. Using equation (8) with $r = a_{\rm d1}$ (equation 9), we find

$$\dot{M}_{\rm visc} \approx \frac{fm_1}{t_{\rm visc}} \approx \frac{fm_1^{3/2} G^{1/2}\alpha_{\rm SS}\theta^2}{2^{1/2} r_1^{3/2}} \approx 91\,{\rm M_\odot\,yr^{-1}}\, f\theta^2\left(\frac{\alpha_{\rm SS}}{0.01}\right)\left(\frac{m_1}{0.3\,{\rm M_\odot}}\right)^{3/2}\left(\frac{r_1}{0.4\,{\rm R_\odot}}\right)^{-3/2}, \tag{28}$$

where $\theta \sim 1$ since the disc initially cannot cool efficiently. This accretion rate is several orders of magnitude higher than the Eddington rate of (C21)

$$\dot{M}_{\rm Edd} \approx 2.7\times 10^{-5}\,{\rm M_\odot\,yr^{-1}}\left(\frac{R_{\rm c}}{0.013\,{\rm R_\odot}}\right). \tag{29}$$

At this stage, radiation is trapped and advected with the flow (e.g. Narayan and Yi 1995; Nordhaus et al. 2011; Shen and Matzner 2014). Most of the mass may be directed into winds/jets (Blandford and Begelman 1999; Armitage and Livio 2000; Hawley and Balbus 2002; Ohsuga et al. 2005), while a fraction accretes onto the central core.

This phase may be sustained by outflows or it may transition into an Eddington-limited phase if the accretion is quenched due to the buildup of gas pressure, which would happen on roughly a dynamical timescale,

$$t_{\rm dyn} \sim \left.\frac{1}{\Omega}\right|_{a_{\rm d1}} \approx \left(\frac{2r_1^3}{Gm_1}\right)^{1/2} \approx 17\,{\rm min}\left(\frac{m_1}{0.3\,{\rm M_\odot}}\right)^{-1/2}\left(\frac{r_1}{0.4\,{\rm R_\odot}}\right)^{3/2}, \tag{30}$$

where we made use of equations (2) and (9). The mass accreted during this time is

$$M_{\rm adv} \approx \dot{M}_{\rm visc} t_{\rm dyn} \approx f\alpha_{\rm SS}\theta^2 m_1 \approx 3\times 10^{-3}\,{\rm M_\odot}\, f\theta^2\left(\frac{\alpha_{\rm SS}}{0.01}\right)\left(\frac{m_1}{0.3\,{\rm M_\odot}}\right), \tag{31}$$

Thus, during the advection-dominated phase, a few ${\rm M_J}$ of material would be deposited into an envelope around the WD (c.f. Nordhaus et al. 2011). Subsequently, the WD may accrete a large fraction of the remaining disc material on the timescale

$$t_{\rm acc} \sim \frac{m_{\rm acc}}{\dot{M}_{\rm Edd}} \sim 4\times 10^3\,{\rm yr}\left(\frac{m_{\rm acc}}{0.1\,{\rm M_\odot}}\right)\left(\frac{R_{\rm c}}{0.0126\,{\rm R_\odot}}\right)^{-1}, \tag{32}$$

where $m_{\rm acc}$ is the mass of accreted material. Remaining disc material would gradually disperse on a timescale of perhaps $\sim$10 Myr, as estimated in Appendix A.

## Appendix E. Powering envelope ejection

The rate of energy release during this advection-dominated phase is roughly given by

$$L_{\rm adv} \approx \frac{GM_{\rm c}\dot{M}_{\rm visc}}{2R_{\rm c}} \approx 2.5\times 10^{44}\,{\rm erg\,s^{-1}}\, f\theta^2\left(\frac{\alpha_{\rm SS}}{0.01}\right)\left(\frac{m_1}{0.3\,{\rm M_\odot}}\right)^{3/2} \times \left(\frac{r_1}{0.4\,{\rm R_\odot}}\right)^{-3/2}\left(\frac{M_{\rm c}}{0.576\,{\rm M_\odot}}\right)\left(\frac{R_{\rm c}}{0.0126\,{\rm R_\odot}}\right)^{-1}, \tag{33}$$

where we have scaled $R_{\rm c}$ to the measured WD radius $R_{\rm wd} = 0.01263 \pm 0.0050\,{\rm R_\odot}$ (Xu et al. 2021). The energy liberated is estimated as

$$E_{\rm adv} \approx L_{\rm adv} t_{\rm dyn} \approx 3\times 10^{47}\,{\rm erg}\, f\theta^2\left(\frac{\alpha_{\rm SS}}{0.01}\right)\left(\frac{m_1}{0.3\,{\rm M_\odot}}\right) \times \left(\frac{M_{\rm c}}{0.576\,{\rm M_\odot}}\right)\left(\frac{R_{\rm c}}{0.0126\,{\rm R_\odot}}\right)^{-1}, \tag{34}$$

which is approximately the same as the initial binding energy of the envelope $E_{\rm bind}$ (Table 1). If about 10% of the original binding energy remains at the time of tidal disruption, then the energy liberated is about an order of magnitude larger than the envelope binding energy. Thus, what remains of the envelope can be rapidly unbound if the energy transfer efficiency of the unbinding process is $\gtrsim 0.1$. Nuclear burning of accreted material may also assist envelope unbinding (Siess and Livio 1999*a*,b; Nordhaus et al. 2011; C21). The disc, on the other hand, is not as susceptible to unbinding both because its binding energy is larger – $E_{\rm d} \sim GM_{\rm c} f m_1/a_{\rm d1} \sim 10^{48}$ erg – and because the released energy may be transported perpendicular to the disc.

If accretion proceeds at the Eddington rate thereafter (equation 32), then energy would be released at the rate

$$L_{\rm Edd} = 7.2 \times 10^{37}\,{\rm erg\,s^{-1}} \left(\frac{M_{\rm c}}{0.576\,{\rm M_\odot}}\right). \tag{35}$$

If, as argued above, $\sim 3 \times 10^{47}$ erg must be released to unbind the envelope (which already factors in the efficiency), we find that the envelope can be ejected in $\sim 10^2$ yr by Eddington-limited accretion alone.

Alternatively, the envelope might be removed before tidal disruption but after the onset of RLOF. In this case, once the envelope is ejected, further orbital tightening would need to be driven by a mechanism other than CE drag. Torques may arise at $a \approx a_{\rm RLOF}$ that lead to orbital decay on timescales of a few hundred orbital periods (c.f. MacLeod et al. 2018). Mass transfer is expected to be unstable for low-mass MS stars (e.g. Stevens et al. 1992; Jones 2020). Thus, even if the envelope is ejected at a separation $a_{\rm d1} < a < a_{\rm RLOF}$, orbital decay down to $a_{\rm d1}$ is still likely to occur. To determine whether this scenario is plausible, we estimate the orbital energy released between $a_{\rm RLOF}$ and $a_{\rm d1}$,

$$-\Delta E_{\rm orb} = \frac{GM_{\rm c} m_1}{2} \left(\frac{1}{a_{\rm d1}} - \frac{1}{a_{\rm RLOF}}\right). \tag{36}$$

From Figure 2a for $m_1 = 0.3\,{\rm M_\odot}$ and the AGB model (Model D, green) with $M_{\rm c} = 0.52\,{\rm M_\odot}$, we find $a_{\rm RLOF} \approx 0.72\,{\rm R_\odot}$ and $a_{\rm d1} \approx 0.44\,{\rm R_\odot}$, which gives $-\Delta E_{\rm orb} \approx 3 \times 10^{47}$ erg for $m_1 = 0.3\,{\rm M_\odot}$, which is the same energy estimated to be released by accretion during the advection-dominated accretion phase (equation 34). As we have already argued, this amount of energy is probably sufficient to unbind the remaining envelope. The value of $\Delta E_{\rm orb}$ is not sensitive to which of the models in Table 1 is adopted for the primary star.

Envelope removal may also be powered by shocking during accretion disc formation (e.g. Rees 1988; Piran et al. 2015; Bonnerot et al. 2021; Ryu et al. 2023; Steinberg and Stone 2024).